\documentclass[a4paper,11pt]{article}
\usepackage{jheppub} 

\makeatletter
\def\@fpheader{}
\makeatother

\usepackage{graphicx}
\usepackage{units}
\usepackage{xcolor}

\newcommand{\ii}{{\rm i}}

\def \epsilon {\varepsilon} 

\def \vec#1{{\boldsymbol{#1}}}

\newcommand{\BR}{\ensuremath{\text{BR}}}

\newcommand{\matrixx}[1]{\begin{pmatrix} #1 \end{pmatrix}} 

\newcommand{\parentheses}[1]{\left( #1 \right)}

\title{Baryon number violation involving tau leptons}

\author{Julian Heeck,}
\affiliation{Department of Physics, University of Virginia,
Charlottesville, Virginia 22904-4714, USA}
\emailAdd{heeck@virginia.edu}

\author{Dima Watkins,}
\emailAdd{bem8mq@virginia.edu}

\abstract{
Baryon number violation is our most sensitive probe of physics beyond the Standard Model, especially through the study of nucleon decays. Angular momentum conservation requires a lepton in the final state of such decays, kinematically restricted to electrons, muons, or neutrinos. We show that operators involving taus, which are at first sight too heavy to play a role in nucleon decays, still lead to clean nucleon decay channels with tau neutrinos. While many of them are already constrained from existing two-body searches such as $p\to \pi^+\nu$, other operators induce many-body decays such as $p \to \eta \pi^{+} \bar\nu_\tau$ and $n\to K^+ \pi^-\nu_\tau$ that have never been searched for.
}

\begin{document}
\maketitle
\flushbottom

\vspace{-3ex}

\section{Introduction}
\label{sec:intro}

The Standard Model (SM) of particle physics predicts baryon number $B$ and lepton number $L$ to be conserved in all interactions at the perturbative level~\cite{FileviezPerez:2022ypk}. Non-perturbative effects generate $\Delta B = \Delta L =3$ processes, but these are suppressed beyond observability at zero temperature~\cite{tHooft:1976rip}.
Physics beyond the SM can in principle violate both $B$ and $L$, famous examples being grand unified theories and supersymmetric SM extensions~\cite{Nath:2006ut}. The experimental signatures are spectacular: atomic matter would decay.
Due to the relative ease in collecting large amounts of matter and observing them over long periods of time, lower limits on proton and neutron decays are unfathomably long, in some cases exceeding the age of our universe by 24 orders of magnitude (i.e.~$\unit[10^{34}]{yr}$~\cite{Super-Kamiokande:2020wjk}).

Any baryon-number-violating proton or neutron decay requires an odd number of leptons in the final state to conserve angular momentum, e.g.~in the form $p\to e^+\pi^0$ or $p\to e^-\mu^+\mu^+$. While electrons, muons, and neutrinos are kinematically allowed final states, taus are roughly twice as heavy as protons and hence cannot be produced. At first sight, this makes tau decays such as $\tau^+\to p \pi^0$ the better $\Delta B$ signature to search for. Alas, taus are rather difficult to produce and detect, rendering these searches~\cite{CLEO:1999emi,Belle:2020lfn} far less sensitive than proton-decay searches.
The crucial observation was already made by Marciano~\cite{Marciano:1994bg} almost three decades ago: any operator or new-physics model that would lead to $\tau^+\to p \pi^0$ would also induce $n \to \pi^0 \bar{\nu}_\tau$ or $p \to \pi^+ \bar{\nu}_\tau$, which are much more sensitive~\cite{Super-Kamiokande:2013rwg} despite the unobservable final-state tau-neutrino.
In the worst-case scenario, one can expect  comparable decay rates,
\begin{align}
\Gamma (n\to \bar{\nu}_\tau \pi^0) \sim \Gamma (\tau^+ \to p \pi^0)\simeq \frac{1}{\unit[10^{33}]{yr}}\,\frac{\BR (\tau^+ \to p \pi^0)}{10^{-53}}  \,,
\label{eq:worst_case}
\end{align}
which would force the $\Delta B$ tau branching ratios at least 40 orders of magnitude below any currently conceivable experimental limits~\cite{Banerjee:2022vdd}.
Here, we quantify this connection more carefully, identify scenarios that violate it and allow for faster tau decays, and emphasize the importance  of neutrino final states in nucleon decays to study tau operators~\cite{Heeck:2019kgr}.

\section{Dimension-six operators}

\begin{table}[tb]
\centering
\begin{tabular}{c c c c} 
 \hline
 field & chirality & generations & $SU(3)_C\times SU(2)_L\times U(1)_Y$ representation  \\ [0.5ex] 
 \hline\hline
 $Q$ & left & 3 & $\left(\vec{3},\vec{2},\tfrac16\right)$\\
 $u$ & right & 3 & $\left(\vec{3},\vec{1},\tfrac23\right)$\\
 $d$ & right & 3 & $\left(\vec{3},\vec{1},-\tfrac13\right)$\\
 $L$ & left & 3 & $\left(\vec{1},\vec{2},-\tfrac12\right)$\\
 $\ell$ & right & 3 & $\left(\vec{1},\vec{1},-1\right)$\\
 $H$ & scalar & 1 & $\left(\vec{1},\vec{2},\tfrac12\right)$\\
 \hline
\end{tabular}
\caption{SM fields and quantum numbers; hypercharge is related to electric charge via $Q = Y + T_3$.}
\label{tab:fields}
\end{table}

The Lagrangian of the Standard Model Effective Field Theory (SMEFT) consists of the usual SM fields from Tab.~\ref{tab:fields} but allows for non-renormalizable higher-dimensional operators, see Ref.~\cite{Isidori:2023pyp} for a recent review; $\Delta B\neq 0$ operators start to appear at operator mass dimension $d=6$~\cite{Weinberg:1979sa}. These $\Delta B=\Delta L =1$ operators can be written as
\begin{align}
\mathcal{L}_{d=6} &= 
y_{abcd}^1 \epsilon^{\alpha\beta\gamma}(\overline{d}^C_{a,\alpha} u_{b,\beta})(\overline{Q}^C_{i,c,\gamma}\epsilon_{ij} L_{j,d}) \\
&\quad +y_{abcd}^2 \epsilon^{\alpha\beta\gamma} \epsilon_{il}\epsilon_{jk} (\overline{Q}^C_{i,a,\alpha} Q_{j,b,\beta})(\overline{Q}^C_{k,c,\gamma}  L_{l,d}) \\
&\quad +y_{abcd}^3 \epsilon^{\alpha\beta\gamma} (\overline{Q}^C_{i,a,\alpha}\epsilon_{ij} Q_{j,b,\beta})(\overline{u}^C_{c,\gamma} \ell_{d}) \\
&\quad +y_{abcd}^4 \epsilon^{\alpha\beta\gamma} (\overline{d}^C_{a,\alpha} u_{b,\beta})(\overline{u}^C_{c,\gamma} \ell_{d})+\text{h.c.} \,,
\label{eq:dequal6}
\end{align}
where  $\alpha,\beta,\gamma$ denote the color, $i,j,k,l$ the $SU(2)_L$, and $a,b,c,d$ the family indices~\cite{Weinberg:1979sa,Wilczek:1979hc,Weinberg:1980bf,Weldon:1980gi,Abbott:1980zj}.
Here, and in the following, the $n$-dimensional Levi--Civita symbols are normalized to $\epsilon_{12\dots n} = +1$.
Operators involving the two or three lightest quarks can be converted to hadron operators using chiral effective field theory~\cite{Claudson:1981gh,Nath:2006ut} and lattice QCD~\cite{Yoo:2021gql}, yielding mass-mixing terms $\bar{p}^C \ell$ and $\bar{n}^C\nu$ as well as interaction terms with mesons.
The Wilson coefficients $y^j$ have mass dimension $-2$ and the first-generation entries are constrained to be $<(\unit[\mathcal{O}(10^{15\text{--}16})]{GeV})^{-2}$ due to the induced two-body nucleon decays such as $p\to e^+\pi^0$~\cite{Heeck:2019kgr,Beneito:2023xbk}.

Operators involving second or third-generation \emph{quarks} do not seem to generate nucleon decays, seeing as charm, bottom, and top quarks are all heavier than the proton. However, since quark flavor is not conserved in the SM, loop amplitudes that lead to nucleon decays can be constructed for any heavy-quark operator~\cite{Hou:2005iu,Dong:2011rh,Gargalionis:2024nij}, or they could proceed through off-shell heavy quarks~\cite{Dong:2011rh,Beneke:2024hox}. Conceptually, it is also difficult to imagine the absence of light-quark operators, since any quark-flavor symmetry has to be broken in nature to comply with the non-diagonal Cabibbo--Kobayashi--Maskawa mixing matrix. For these reasons, we will mostly restrict ourselves to first-generation quarks in the following, except for operators that would vanish in this one-generational limit.

\begin{figure}[tb]
    \centering
    \includegraphics[width=0.85\textwidth]{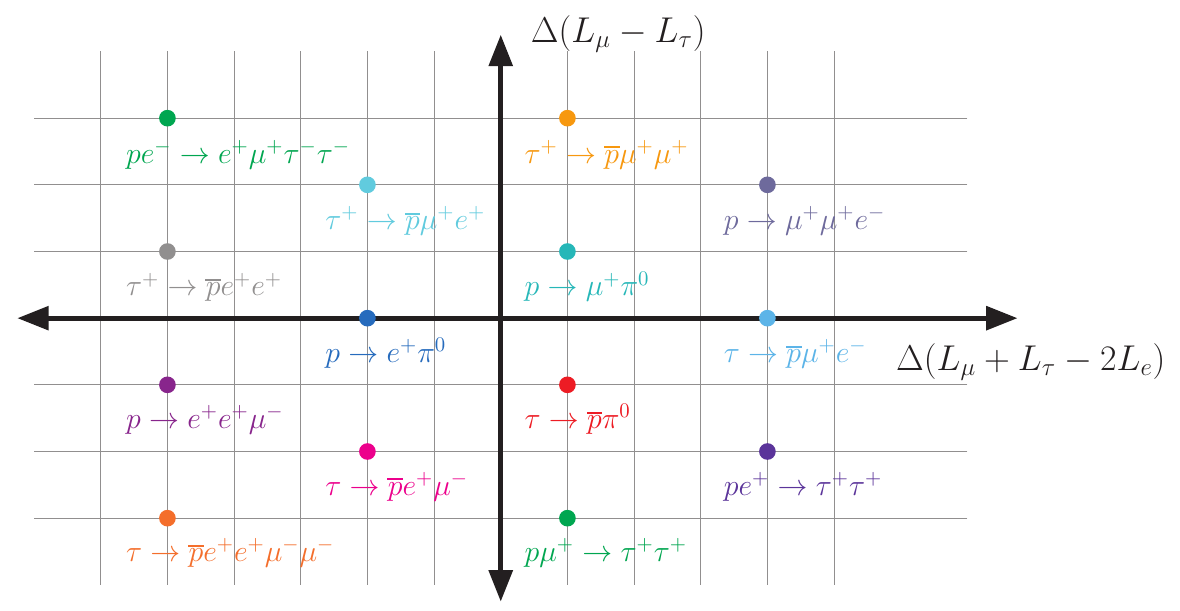}
    \caption{Landscape of $\Delta B = \Delta L = 1$ operators organized by their lepton-flavor structure. We only show one example process for each group, others are implied.
    Experimental limits for $p\to e^+ \pi^0, \mu^+\pi^0$~\cite{Super-Kamiokande:2020wjk} and $p\to e^+ e^+\mu^-, \mu^+\mu^+ e^-$~\cite{Super-Kamiokande:2020tor} come from Super-Kamiokande, for $\tau\to p \ell \ell$ (except for the missing channel $\tau^+\to \bar{p} \mu^+ e^+$) from Belle~\cite{Belle:2020lfn}, and for $\tau\to \bar{p}\pi^0$ there exist 25-year-old limits from CLEO~\cite{CLEO:1999emi}.
    }
    \label{fig:LFV_grid_B1L1}
\end{figure}

The same argument does not apply to the \emph{lepton} generations though. The three individual lepton numbers, electron, muon, and tau, are conserved in the SM, so operators with $\Delta L_\tau=1$, for example, will never lead to $\Delta L_\tau = 0$ processes.
Even the observation of neutrino oscillations does not quantitatively change this conclusion: the continued absence of any charged-lepton flavor violation~\cite{Davidson:2022jai} can be taken as an indication that lepton flavor is only violated through neutrino masses. If any and all lepton flavor violation is suppressed by neutrino masses, the effects are near impossible to observe and render lepton flavor an incredibly good approximate symmetry in the charged lepton sector. 
SMEFT operators can then be organized according to their quantum numbers under the global SM symmetry group $U(1)_{B+L}\times U(1)_{B-L}\times U(1)_{L_\mu-L_\tau}\times U(1)_{L_\mu + L_\tau - 2L_e}$, seeing as these symmetries are either extremely good approximate or even exact symmetries~\cite{Heeck:2014zfa,Heeck:2016xwg}. An example is shown in Fig.~\ref{fig:LFV_grid_B1L1}, organizing all $\Delta B = \Delta L = 1$ operators/processes by their lepton-flavor content. Only the three groups closest to the origin ($p\to e^+\pi^0$, $p\to \mu^+\pi^0$, and $\tau\to \bar{p} \pi^0$) arise at $d=6$, the others require $d\geq 10$.
It is easy to impose lepton numbers as global or even local $U(1)$ symmetries, broken only in the neutrino sector~\cite{Heeck:2016xwg}, that forbid all but one group in  Fig.~\ref{fig:LFV_grid_B1L1}.
Similarly, we can easily construct models in which baryon number is only broken together with some linear combination of lepton flavor~\cite{Hambye:2017qix,Heeck:2019kgr}.\footnote{This example is already realized in the SM, where $\Delta B/3 = \Delta L_e = \Delta L_\mu = \Delta L_\tau = 1$ via instantons.}
This is sufficient motivation for a dedicated study of $\Delta B$ operators involving taus, which are usually ignored due to the kinematics but could well be the only baryon-number violating processes in nature. For example, if we impose $U(1)_{B - L_\tau}$ on the $d=6$ operators from Eq.~\eqref{eq:dequal6}, we are left with tau operators.

Using the Chiral Perturbation Theory (ChPT) framework from Refs.~\cite{Claudson:1981gh,Nath:2006ut}, summarized in App.~\ref{sec:chiralpt_appendix}, we can calculate the dominant baryon-number-violating tau decays induced by the four $d=6$ operators:
\begin{align}
    \Gamma (\tau^+\to p\pi^0) & \simeq \tfrac12 \Gamma (\tau^+\to n\pi^+)\nonumber \\
& \simeq \frac{(|g_L^\pi|^2 + |g_R^\pi|^2) \parentheses{m_{\tau}^{2} + m_{p}^{2} - m_{\pi}^{2}}  - 4\, \text{Re} \left( g_R^\pi g_L^{\pi *}\right) m_\tau m_p}{32 \pi m_{\tau}^{3}/\sqrt{\lambda\parentheses{m_{\tau}^{2}, m_{p}^{2}, m_{\pi}^{2}}}} \,,\\
\Gamma (\tau^+\to p\eta) & \simeq   \frac{(|g_L^\eta|^2 + |g_R^\eta|^2) \parentheses{m_{\tau}^{2} + m_{p}^{2} - m_{\eta}^{2}}  - 4\, \text{Re} \left( g_R^\eta g_L^{\eta *}\right) m_\tau m_p }{32 \pi m_{\tau}^{3}/\sqrt{\lambda\parentheses{m_{\tau}^{2}, m_{p}^{2}, m_{\eta}^{2}}}} \,,
\end{align}
where $\lambda (x,y,z) \equiv x^2 + y^2 + z^2 - 2 x y - 2 x z - 2 y z$ is the K\"all\'en polynomial,
\begin{align}
g_L^\pi &\equiv \ii \frac{c_L \left[ (D+F+1)m_p^2 + (D+F-1)m_\tau^2 \right]+ 2 c_R (D+F) m_p m_\tau}{\sqrt{2} f_\pi (m_p^2-m_\tau^2)} \,,\\
g_L^\eta &\equiv \ii \frac{c_R' \left(m_p^2 - m_\tau^2 \right)+ 2 (D-3 F) m_\tau (c_L m_p + c_R m_\tau)}{\sqrt{6} f_\pi (m_p^2-m_\tau^2)} \,,
\end{align}
and $g_R^{\pi,\eta}\equiv g_L^{\pi,\eta}(c_L\leftrightarrow -c_R; c_L'\leftrightarrow -c_R')$, 
and the relevant linear combinations of Wilson coefficients
\begin{align}
    c_L &\equiv \alpha C_{RL}^{\tau} + \beta C_{LL}^{\tau} \,, & c_L' &\equiv \alpha \left( D -3F+1\right) C_{RL}^{\tau} + \beta \left(D - 3F -3\right)C_{LL}^{\tau} \,,\\
    c_{R} &\equiv \alpha C_{LR}^{\tau} + \beta C_{RR}^{\tau}\,, & c_{R}' &\equiv \alpha \left(D -3F + 1\right) C_{LR}^{\tau} + \beta \left(D - 3F -3\right) C_{RR}^{\tau} \,.
\end{align}
Here and below, we kept the notation for the Wilson coefficients from Refs.~\cite{Claudson:1981gh,Nath:2006ut} for relative brevity, but use the relationship $C^{\nu_\tau}_{AB} = - C^\tau_{AB}$ that arises from the $SU(2)_L$ symmetry of our SMEFT operators; the mapping is given in Tab.~\ref{tab:mapping}.
For numerical evaluations, we use $\alpha \simeq -\beta \simeq -\unit[0.013]{GeV^3}$~\cite{Yoo:2021gql} from lattice QCD, $D=0.62$, $F=0.44$ from a chiral-perturbation-theory analysis of hyperon decays~\cite{Ledwig:2014rfa}, and all other parameters from the Particle Data Group~\cite{ParticleDataGroup:2022pth}, in particular $f_\pi \simeq \unit[130]{MeV}$ in our convention.

\begin{table}[tb]
    \centering
    \begin{tabular}{ c | c }
    \hline
        Chiral Perturbation Theory Coefficient & SMEFT Wilson Coefficient \\
    \hline    \hline
        $C_{RL}^{\nu_{\tau}}$ & $- y_{1113}^{1}$  \\
    \hline
        $C_{LL}^{\nu_{\tau}}$ & $- y_{1113}^{2}$ \\
    \hline
        $C_{RL}^{\tau}$ & $ y_{1113}^{1}$ \\
    \hline
        $C_{LR}^{\tau}$ & $-2 y_{1113}^{3}$ \\
    \hline
        $C_
        {LL}^{\tau}$ & $y_{1113}^{2}$ \\
    \hline
        $C_{RR}^{\tau}$ & $y_{1113}^{4}$ \\
    \hline
    \end{tabular}
    \caption{ChPT coefficient~\cite{Nath:2006ut} relationship to our Wilson coefficients from Eq.~\eqref{eq:dequal6}; all carry units of $  \unit{GeV}^{-2}$.}
    \label{tab:mapping}
\end{table}

These $\Delta B = \Delta L_\tau = 1$ tau decays are constrained by CLEO to rates $\Gamma (\tau^+\to p\pi^0)< 1.5\times 10^{-5}\,\Gamma_\tau \simeq \unit[3.4\times 10^{-17}]{GeV}\simeq (\unit[2\times 10^{-8}]{s})^{-1}$~\cite{CLEO:1999emi}, and similarly for the $\eta$ mode. Judging by their results in Ref.~\cite{Belle:2020lfn}, Belle could improve these bounds by three orders of magnitude with their large existing data set, and Belle II could eventually improve them by another two orders of magnitude~\cite{Belle-II:2022cgf,Banerjee:2022vdd}.
Translated into upper bounds on the SMEFT Wilson coefficients, we obtain limits of order $|C| < (\unit[1.2]{TeV})^{-2}$ for the pion mode and $|C| < (\unit[1.2]{TeV})^{-2}$ or $|C| < (\unit[1.6]{TeV})^{-2}$  for the eta channels 
from the CLEO constraints, more detailed in Tab.~\ref{tab:limits}. These limits probe viable SMEFT parameter space since the EFT scale is clearly pushed above the electroweak scale.
Belle II could conceivably improve the reach to $|C| < (\unit[20]{TeV})^{-2}$. 
Still, the limits are nowhere near typical proton-decay scales, making it crucial to evaluate nucleon-decay channels mediated by the same operators.

\begin{table}[tb]
    \centering
    \begin{tabular}{ c | c | c | c | c  }
    \hline
        Process & 
        $C_{RL}^{\tau}$ $[\unit{GeV^{-2}}]$ & 
        $C_{LL}^{\tau}$ $[\unit{GeV^{-2}}]$ & 
        $C_{LR}^{\tau}$ $[\unit{GeV^{-2}}]$ &
        $C_{RR}^{\tau}$ $[\unit{GeV^{-2}}]$ \\
    \hline\hline
         $\tau^{+} \to p^{+} \pi^{0}$ & 
         $6.9 \times 10^{-7}$ &
         $6.9 \times 10^{-7}$ &
         $6.9 \times 10^{-7}$ &
         $6.9 \times 10^{-7}$ \\
    \hline
         $\tau^{+} \to p^{+} \eta$ & 
         $7.5 \times 10^{-7}$ &
         $3.9 \times 10^{-7}$ & 
         $7.5 \times 10^{-7}$ & 
         $3.9 \times 10^{-7}$\\
    \hline\hline
         $n \to \overline{\nu}_{\tau} \pi^{0}$ &
         $3.1 \times 10^{-31}$ &
         $3.1 \times 10^{-31}$ &
         N.A. &
         N.A. \\
    \hline
         $n \to \overline{\nu}_{\tau} \eta$
         & $1.5 \times 10^{-29}$  &
         $1.2 \times 10^{-30}$
         & N.A.
         & N.A. \\
    \hline
         $p \to \overline{\nu}_{\tau} \pi^{+} $ &
         $3.7 \times 10^{-31}$ 
         &
         $3.7 \times 10^{-31}$ &
         $3.8 \times 10^{-24}$ &
         $3.8 \times 10^{-24}$ \\
    \hline
         $p \to e^{+} \nu_{e} \overline{\nu}_{\tau}$ &
         $1.6 \times 10^{-23}$ &
         $1.6 \times 10^{-23}$ &
         $8.5 \times 10^{-24}$ &
         $8.5 \times 10^{-24}$ \\
    \hline
    \end{tabular}
    \caption{Current upper limits on $\Delta B = \Delta L_\tau = 1$ $d=6$ Wilson coefficients from the processes in the first column, assuming one non-vanishing operator at a time. See text for details. N.A. means the process arises at loop level, not calculated here. 
}
    \label{tab:limits}
\end{table}

\begin{figure}[tb]
    \centering
    \includegraphics[width=0.999\textwidth]{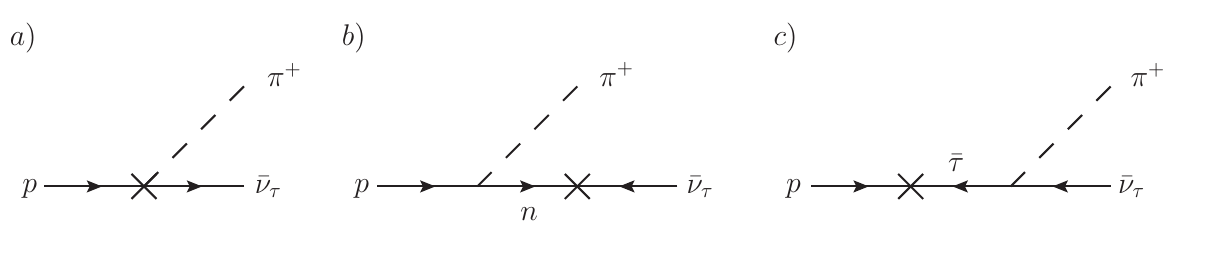}
    \caption{Proton decay into $\pi^+\bar{\nu}_\tau$ through various dimension-six operators, indicated with a cross.
    }
    \label{fig:proton_decay_through_tauon}
\end{figure}

$\mathcal{L}_{d=6}$ operators involving \textit{left-handed} taus unavoidably come with $\nu_\tau$ operators that directly lead to $p \to \bar \nu_\tau \pi^+$  [see Fig.~\ref{fig:proton_decay_through_tauon}a) and b)], $n\to \pi^0\bar{\nu}_\tau$ and $n \to \bar \nu_\tau \eta$, and we generically expect the relationship of Eq.~\eqref{eq:worst_case} to hold.
This estimate is confirmed with the more carefully calculated expressions for the nucleon decay rates
\begin{align}
 \Gamma (p \to \bar \nu_\tau \pi^+) & \simeq 2 \Gamma (n \to \bar \nu_\tau \pi^0)  \simeq  \frac{1}{32 \pi f_{\pi}^{2}} \frac{(m_{n}^{2} - m_{\pi}^2)^{2}}{m_{n}^{3}} (1 + D + F)^{2} |c_{L}|^{2}  \,,\\ 
\Gamma (n \to \bar \nu_\tau \eta) & \simeq  \frac{1}{192 \pi f_{\pi}^{2}} \frac{(m_{n}^{2} - m_{\eta}^{2})^{2}}{m_{n}^{3}}|c_{L}'|^{2}\,,
\end{align}
currently constrained by Super-K and IMB-3 to $\Gamma (p\to \bar \nu_\tau \pi^+) < (\unit[3.9\times 10^{32}]{yr})^{-1}$~\cite{Super-Kamiokande:2013rwg}, $\Gamma (n\to \bar \nu_\tau \pi^0) < (\unit[1.1\times 10^{33}]{yr})^{-1}$~\cite{Super-Kamiokande:2013rwg}, and $\Gamma (n \to \bar \nu_\tau \eta) < (\unit[1.6\times 10^{32}]{yr})^{-1}$~\cite{McGrew:1999nd}, respectively!
This forces the Wilson coefficients at least 24  orders of magnitude down compared to the tau limits, assuming one non-zero Wilson coefficient at a time, see Tab.~\ref{tab:limits}. 
Operators involving \textit{right-handed} taus do not directly come with $\nu_\tau$ operators and thus seemingly circumvent the dangerous nucleon decays into tau neutrinos; however, even these lead to $p\to \pi^+\bar{\nu}_\tau$ through an off-shell tau [see Fig.~\ref{fig:proton_decay_through_tauon}c)], as pointed out long ago by Marciano~\cite{Marciano:1994bg}. The off-shell tau propagator and required mass flip do not cause any suppression since $m_\tau \sim m_p$, but the off-shell tau \textit{decay} comes with a $G_F$ suppression, which gives roughly  $\Gamma (p\to \bar{\nu}_\tau \pi^+) \sim (G_F f_\pi)^2 \Gamma (\tau^+ \to p \pi^0)\simeq \BR (\tau^+ \to p \pi^0)(\unit[10^{-6}]{yr})^{-1}$, still forcing $\BR (\tau^+ \to p \pi^0)$ below $10^{-40}$. The full expression reads
\begin{align}
\Gamma (p \to \bar \nu_\tau \pi^+) & \simeq  \frac{1}{32 \pi f_{\pi}^2} \frac{(m_{p}^{2} - m_{\pi}^{2})^{2}}{m_{p}^{3}} \left| (1+D+F) c_L - \frac{\sqrt{2} c_1 f_\pi^2 G_F m_p m_\tau}{m_p^2-m_\tau^2} c_R\right|^2 ,
\end{align}
with Cabibbo cosine $c_1\equiv \cos\theta_c\simeq 0.97$.

The $n\to \bar{\nu}_\tau$ channels give limits on the left-handed Wilson coefficients between $(\unit[3\times 10^{14}]{GeV})^{-2}$ and $(\unit[2\times 10^{15}]{GeV})^{-2}$ (see Tab.~\ref{tab:limits}), probing scales at least eleven orders of magnitude above current tau decay limits and ten orders of magnitude above future ones. 
Despite the relative suppression by $G_F f_\pi^2\sim 10^{-7}$ for the \emph{right-handed} tau operators, the resulting limits of order $(\unit[5\times 10^{11}]{GeV})^{-2}$  still far exceed any conceivable tau-decay limits~\cite{Hou:2005iu}.
The proton--tau mixing operators also induce the fully leptonic three-body decays
\begin{align}
\Gamma (p \to e^+ \nu_e \bar \nu_\tau) & \simeq \Gamma (p \to \mu^+ \nu_\mu \bar  \nu_\tau) \simeq \frac{G_{F}^{2} m_{p}^{5}}{192 \pi^{3} \left(m_{\tau}^{2} - m_{p}^{2}\right)^{2}} |c_{L} m_{p} - c_{R}m_{\tau}|^{2}\,,
\end{align}
with the same charged-lepton energy spectrum as in tau decays. A dedicated Super-Kamiokande search for precisely such three-body decays has led to a limit $\Gamma^{-1} (p \to \ell^+ \nu_\ell \bar \nu_\tau) > \unit[2\times 10^{32}]{yr}$~\cite{Super-Kamiokande:2014pqx}, which gives Wilson coefficient constraints of the same order of magnitude as the two-body decays $p\to \pi^+ \bar{\nu}_\tau$, see Tab.~\ref{tab:limits}.
At even higher order in perturbation theory, $p$ can decay into many other final states through an off-shell tau, including the four-body decay $p\to  e^{+} \nu_{e}\overline{\nu}_{\tau} \pi^{0}$ 
recently carefully calculated in Ref.~\cite{Crivellin:2023ter}, although these are far less important than $p\to \bar \nu_\tau \pi^+$, both due to phase-space suppression and lack of dedicated searches.

\begin{figure}[tb]
    \centering
    \includegraphics[width = 0.6\textwidth]{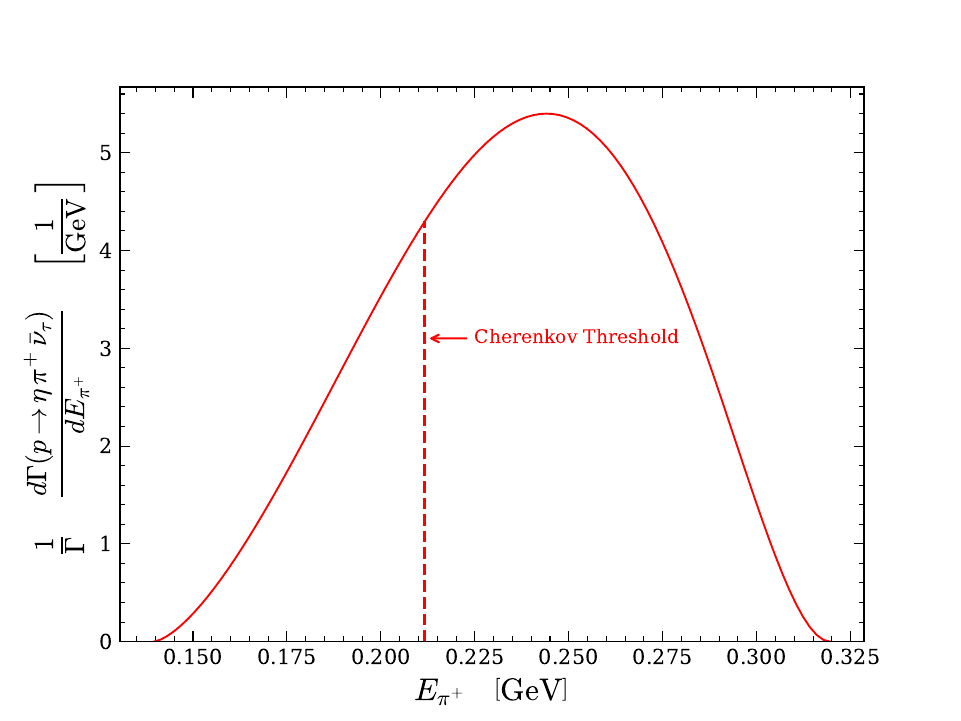}
    \caption{The normalized energy spectrum of the final-state $\pi^{+}$ in the $p \to  \eta \pi^{+} \bar\nu_{\tau}$ mode; 73\% of the pions are above Cherenkov threshold in Super-Kamiokande. 
    }
    \label{fig:pi_plus_spectrum}
\end{figure}

Importantly, all nucleon decays calculated so far only probe three linear combinations of the four $d=6$ Wilson coefficients, being notably independent of $c_R'$. Setting $c_L = c_R = c_L' = 0$ -- which leaves the operator $(\overline{d}^C u)(\overline{u}^C P_R \tau)\propto \eta \,\overline{p}^C P_R \tau$ -- would then seemingly allow for a large $\tau^+\to p\eta$ decay rate without any competing nucleon decays.\footnote{The importance of $\eta$ modes to probe this flat direction in the electron and muon cases has been pointed out recently in Ref.~\cite{Beneito:2023xbk}. } Even the vector-meson final state $p\to \rho^+ \bar \nu_\tau$ vanishes in that limit~\cite{Kaymakcalan:1983uc}.
Alas, the underlying operator of course still generates some form of nucleon decay, for example the three-body proton decay $p\to \eta \pi^+\bar{\nu}_\tau$.
Compared to the two-body tau decay $\tau^+\to p\eta$, this proton decay rate is suppressed by a  phase-space factor $m_p^2/(32\pi^2m_\tau^2)$ times partial phase-space closure from the large $\eta$ mass, and of course still the $G_F^2$ suppression from the off-shell tau decay.
An analytical form for the decay rate can be obtained in the massless pion limit:
\begin{align}
    \Gamma \parentheses{p \to \eta \pi^{+} \bar\nu_\tau} & = 
\frac{G_{F}^{2} c_{1}^{2} |c_{R}'|^{2} m_{p}^{5}}{12288 \pi^{3} m_{\tau}^{2}} \left[ 1 + \frac{44}{3} \parentheses{\frac{m_{\eta}}{m_{p}}}^{2} - 12 \parentheses{\frac{m_{\eta}}{m_{p}}}^{4} + 4 \parentheses{\frac{m_{\eta}}{m_{p}}}^{6} \right. \nonumber\\ 
    & \qquad\qquad\left.+ \frac{1}{3} \parentheses{\frac{m_{\eta}}{m_{p}}}^{8}- 8 \parentheses{\frac{m_{\eta}}{m_{p}}}^{2}\parentheses{2 + 3 \parentheses{\frac{m_{\eta}}{m_{p}}}^{2}} \log \parentheses{\frac{m_{\eta}}{m_{p}}} \right].
\end{align}
Keeping the pion mass suppresses this roughly by a factor of two:
\begin{align}
   \Gamma \parentheses{p \to \eta \pi^{+} \bar\nu_\tau} \simeq \parentheses{\unit[5.6 \times 10^{-19} ]{GeV}^{-1}}\, |c_R'|^2 \simeq \frac{1}{\unit[200]{yr}} \left(\frac{\BR (\tau \to p\eta)}{8.9\times 10^{-6}}\right) .
   \label{eq:p_to_eta_pi_nu}
\end{align}
Even though no dedicated exclusive search for the three-body decay $p \to \eta \pi^{+} \bar\nu_\tau$ exists, ancient inclusive limits should exclude lifetimes below $\unit[10^{30}]{yr}$~\cite{Heeck:2019kgr} and illustrate once again the disparity between $\Delta B$ searches in nucleons and taus, forcing the $\Delta B$ tau branching ratio into $p \eta$ below $10^{-30}$. (Notice though the more than 20 orders of magnitude gain compared to the generic left-handed operator analysis!)  We encourage our experimental colleagues from Super-Kamiokande to perform a dedicated search for $p \to \eta \pi^{+} \bar\nu_\tau$ to improve this limit by orders of magnitude, seeing as this mode is complementary to the typically considered two-body decay modes. We provide the energy spectrum for $\pi^+$ in Fig.~\ref{fig:pi_plus_spectrum}; a fraction of $73\%$ are above the Cherenkov threshold, neglecting the possible slowdown of protons decaying inside the oxygen nucleus. The $\eta$ 
has a typical momentum of $\unit[0.2]{GeV}$ and  
decays dominantly to two photons or three pions ($3\pi^0$ or $\pi^0 \pi^+\pi^-$), leading to a busy signature that might benefit from DUNE's tracking detector.

As an aside, the same cancellation can be employed with similar results for other flavor indices. For example, non-zero $y^{3,4}_{2113}$ induce a mixing of $\Sigma^-$ with the right-handed tau, leading to $\tau\to \Sigma^-\pi^0,\, \Sigma^0\pi^-$ as well as the competing decay $p\to K^0 \pi^+ \bar\nu_\tau$ through off-shell $\tau$. However, we can adjust the two Wilson coefficients to cancel in the proton-decay amplitude without affecting the tau decay. 
This forbids nucleon decays to order $f_\pi^{-1}$ in tree-level strangeness-conserving ChPT  while allowing for two-body $\Delta B$ tau decays, albeit in channels that have never been investigated.
Still, at order $f_\pi^{-2}$ we find tree-level proton decays $p\to K^+\pi^- \pi^+ \bar{\nu}_\tau$ and $p\to K^0\pi^0 \pi^+ \bar{\nu}_\tau$ that survive the Wilson-coefficient cancellation; despite the $f_\pi^{-2}$, $G_F$ and phase-space suppression -- and the fact that only weak inclusive limits for this decay exist -- this will beat any conceivable tau decays.

This concludes our discussion of $d=6$ operators. Assuming one non-zero operator at  a time, existing limits on two-body nucleon decays far outperform even optimistic $\Delta B$ taus decays. This conclusion can be softened a bit by allowing for cancellations between operators, which can relegate nucleon decays to overlooked three- or four-body channels, but even they are indirectly constrained well enough to uphold the above conclusion.

\section{Dimension-seven operators}

Restricting ourselves to non-derivative operators for simplicity~\cite{Weinberg:1980bf}, there are four independent baryon-number-violating operators at $d=7$~\cite{Lehman:2014jma,Liao:2016hru}:
\begin{align}
\mathcal{L}_{d=7} &= z_{abcd}^1 \epsilon^{\alpha\beta\gamma} (\overline{Q}^C_{i,a,\alpha} Q_{j,b,\beta})(\overline{L}_{i,c} d_{d,\gamma}) H_j^*
 \\
&\quad +z_{abcd}^2 \epsilon^{\alpha\beta\gamma}\epsilon_{ij} (\overline{u}^C_{a,\alpha} d_{b,\beta})(\overline{L}_{i,c} d_{d,\gamma}) H_j^*\\
&\quad +z_{abcd}^3 \epsilon^{\alpha\beta\gamma}  (\overline{d}^C_{a,\alpha} d_{b,\beta})(\overline{\ell}_{c}  Q_{i,d,\gamma}) H_i^* \\
&\quad +z_{abcd}^4 \epsilon^{\alpha\beta\gamma} (\overline{d}^C_{a,\alpha} d_{b,\beta})(\overline{L}_{c,i} d_{d,\gamma}) H_i +\text{h.c.} 
\label{eq:dequal7}
\end{align}
The Wilson coefficients $z^j$ now have mass dimension $-3$.
Upon electroweak symmetry breaking, $H\to (0,v/\sqrt{2})$, with $v\simeq\unit[246]{GeV}$, these give $\Delta B=-\Delta L =1$~\cite{Weinberg:1980bf}  four-fermion operators that can be translated to hadronic operators using the ChPT from App.~\ref{sec:chiralpt_appendix}.

$z^1$ and $z^2$ boil down to $ud d\bar{\nu}_\tau $, and thus $n\to \nu_\tau \pi^0$ just like the $d=6$ operators with simple renaming $\bar{\nu}_\tau\to \nu_\tau$ and
\begin{align}
y^1 \to \frac{v}{\sqrt{2}} z^1\,, \qquad
y^2 \to \frac{v}{\sqrt{2}} z^2\,.
\end{align}
This then gives limits $|z^{1,2}_{1131}| < (\unit[10^{11}]{GeV})^{-3}$ and hopelessly suppressed $\Delta B$ tau decay rates. The other two operators give $d_R d_R d_{L,R} \bar{\tau}$, which vanish if all $d$ quarks are from the same generation. The leading operator is then $d_R s_R d_{L,R} \bar{\tau}$, which leads to $p\to \pi^+ K^+ \ell$ or $n\to K^+ \ell$~\cite{Weinberg:1980bf} at tree level; however, for the charged tau this is kinematically forbidden, forcing us to go through an off-shell tau to $p\to \pi^+ K^+ \pi^-\nu_\tau$  or $n\to K^+ \pi^-\nu_\tau$. 
This competes with the two-body tau decay $\tau\to \Lambda^0 \pi^-$ constrained by Belle~\cite{Belle:2005exq}: 
\begin{align}
    \Gamma\parentheses{n\to K^+ \pi^-\nu_\tau}\simeq  \left(\frac{\BR (\tau \to \Lambda^0 \pi^-)}{7.2\times 10^{-8}}\right) \begin{cases}
    \frac{1}{\unit[1.3\times 10^5]{yr}} \,, & z_3 = 0\,,\\
    \frac{1}{\unit[7\times 10^3]{yr}} \,, & z_4 = 0\,.\\
    \end{cases}
\end{align}
No experimental constraints exist for the neutron decay, except for ancient inclusive limits~\cite{Heeck:2019kgr}, which push $\BR (\tau \to \Lambda^0 \pi^-)$ below $10^{-30}$, analogous to Eq.~\eqref{eq:p_to_eta_pi_nu}.
At \emph{loop} level in a UV-complete realization of these operators, we can turn the $dsd \bar{\tau}$ operators into $d s u \bar{\nu}_\tau$ operators, which then give the cleaner two-body decay $p\to  K^+ \nu_\tau$, constrained to $\Gamma^{-1} (p \to K^+ \nu) > \unit[6\times 10^{33}]{yr}$~\cite{Super-Kamiokande:2014otb}. However, the loop amplitude requires mass flips from the tau and one of the quarks, leading to an amplitude suppression by $y_\tau y_d/(16\pi^2)\sim 2\times 10^{-9}$ and limits of order $|z| \sim (\unit[10^{8}]{GeV})^{-3}$. A dedicated search for $n\to K^+ \pi^-\nu_\tau$ could lead to more stringent and reliable results.

\begin{figure}[tb]
    \centering
    \includegraphics[width=0.4\textwidth]{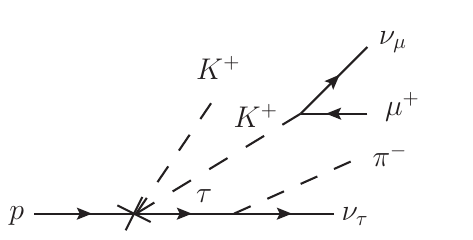}
    \caption{Proton decay into $K^+ \mu^+ \nu_\mu \pi^-\nu_\tau$ through the dimension-seven operator $ds \bar{L}_\tau s H$, indicated with a cross.
    }
    \label{fig:proton_decay_through_tauon3}
\end{figure}

Let us give one more example of how far one would have to suppress nucleon decays to allow for testable $\Delta B=1$ tau decays. We take $z^4_{1232}$, i.e.~a $dss$ operator, which induces a mixing of $\tau^-$ with $\Xi^-$. This still allows for a kinematically allowed two-body tau decay $\tau\to \Xi \pi$,\footnote{Isospin gives approximately $ \Gamma(\tau\to \Xi^0 \pi^-)\simeq 2\, \Gamma(\tau\to \Xi^- \pi^0)$.} but the double strangeness severely suppresses nucleon decays. Only at order $f_\pi^{-2}$ do we find an operator $\bar{\tau} P_R p K^- K^-\, \beta v z^4_{1232}/(\sqrt{2} f_\pi^2)$ that allows for proton decay with emission of tau and two kaons. The tau is necessarily off-shell just like in previous examples, but now even one of the kaons needs to be off-shell too! This is then a doubly $G_F$ suppressed five-body proton decay,  illustrated in Fig.~\ref{fig:proton_decay_through_tauon3}, estimated to
\begin{align}
    \Gamma\parentheses{p\to K^+ \mu^+ \nu_\mu \pi^-\nu_\tau}\simeq  \frac{1}{\unit[\mathcal{O}(10^{28})]{yr}}  \left(\frac{\BR (\tau \to \Xi^- \pi^0)}{ 10^{-8}}\right) ,
\end{align}
where we approximated the amplitude  -- also including a factor $m_\pi^2 m_\mu/(m_\tau^2 m_K^2)$ from the derivative $K$ and $\pi$ vertices and $K$ and $\tau$ propagators -- as constant but included kaon, pion and muon mass in the phase space integration. Still, even with this immense suppression of nucleon decay due to $G_F^2$ and the five-body phase space we are falling short of realistic viable lifetimes.
Nevertheless, we urge our experimental colleagues to investigate $\Delta B$ tau decays into hyperons as they have the most suppressed associated nucleon decays.

\section{Operators of higher mass dimension}

$\Delta B$ operators at $d=8$ have $\Delta B = \Delta L =1$, just like the $d=6$ ones. A linearly independent basis of $\Delta B$ operators can be found in Refs.~\cite{Murphy:2020rsh,Li:2020gnx}. By and large, the phenomenology is similar to the $d=6$ operators; one crucial difference is the occurrence of operators involving charm quarks due to anti-symmetry. These either require a hadron framework beyond our simple chiral perturbation theory or the inclusion of loops~\cite{Gargalionis:2024nij}, which will once again render the phenomenology similar if not identical to $d=6$.

Dimension-nine operators involving taus and first-generation up quarks necessarily violate $\Delta B = -\Delta L = 1$, just like the $d=7$ operators~\cite{Heeck:2019kgr}, and indeed many operators share features with those from $d=7$. The obvious exceptions are the $d=9$ operators involving three quarks, one lepton, and two anti-leptons, already discussed in Ref.~\cite{Hambye:2017qix} because of their potentially unique lepton-flavor properties.
As an example, consider the operator $ddd\ell\bar\ell\bar\ell/\Lambda^5$; antisymmetry enforces once of the down quarks to be a strange, and two different flavors of antileptons, e.g. $dsd \tau\bar{e}\bar{\mu}/\Lambda^5$. This gives $\tau\to \Sigma^+e\mu$, competing with $n \to K^+ e \mu \pi^+\bar{\nu}_\tau$ through an off-shell tau, neither of which has been searched for of course.

Dimension-ten operators have been discussed in Refs.~\cite{Hambye:2017qix,Heeck:2019kgr}, let us focus on one example operator here, $QQ u\ell \bar{\ell} L H/\Lambda^6$. Upon electroweak symmetry breaking, this can give $u_L d_L u_R \overline{\tau_R} \ell_\alpha\ell_\beta\, v/\Lambda^6$; if $\alpha$ and $\beta$ correspond to electrons or muons, this operator generates the tau decays $\tau^+ \to \bar{p} e^+_\alpha e^+_\beta$ shown in Fig.~\ref{fig:LFV_grid_B1L1}, recently searched for in Belle~\cite{Belle:2020lfn}. The same operator also generates $p\to e^+_\alpha e^+_\beta \pi^- \nu_\tau$ through the off-shell tau:
\begin{equation}
    \Gamma\parentheses{p\to e^+_\alpha e^+_\beta \pi^- \nu_\tau}\simeq \frac{1}{\unit[2\times 10^4]{yr}} \left(\frac{\BR (\tau^- \to p e^-_\alpha e^-_\beta)}{3\times 10^{-8}}\right) .
\end{equation}
Despite the four-body vs three-body phase space suppression and the fact that no exclusive limits on this channel exist, the proton decay will enforce a sufficient limit on the Wilson coefficient as to make the $\Delta B$ tau decay completely unobservable.

\section{Conclusions}
\label{sec:conclusions}

Nobody knows if the conservation of baryon number in the SM is a fundamental feature or a happy little accident that doesn't survive SM extensions. On the one hand, life crucially relies on sufficiently long-lived nuclei; on the other hand, the apparent asymmetry of matter over antimatter in our universe seems to suggest that baryon number might be violated. This is ultimately an experimental question that will hopefully be answered eventually, for example by observing nucleon decays in detectors such as Super-Kamiokande or DUNE.
No positive signal for baryon number violation has been observed yet despite decade-long efforts, which could, however, simply mean that we are not looking in the right spots. From what little we know about the flavor structure of the SM, it could well be that baryon number is mainly or even only broken together with tau number, which naively changes the expected signatures since taus are heavier than nucleons. Indeed, searches for baryon-number-violating tau decays such as $\tau\to \bar{p} \pi^0$ have been performed. As pointed out long ago by Marciano, however, the underlying new physics will also generate nucleon decays such as $p\to \pi^+\bar{\nu}_\tau$, which is far more sensitive.
Here, we have studied the relationship between $\Delta B$ nucleon and tau decays quantitatively for a large number of new-physics operators to confirm Marciano's observation and scrutinize loopholes.
As expected, we find that any operator that leads to $\Delta B=1$ tau decays also leads to nucleon decays, the tau flavor being carried away by tau neutrinos. However, it is not difficult to find examples in which the nucleon only decays in channels that have never been explicitly searched for, which significantly softens the relationship but does not practically change it, since even old weak inclusive limits are sufficient to beat tau limits.
We stress that this conclusion should in no way discourage anyone from searching for $\Delta B$ tau decays; if anything, this is meant to encourage broadening searches for nucleon decays, either by  going beyond two-body final states, e.g.~$p \to \eta \pi^{+} \bar\nu_\tau$ and $n\to K^+ \pi^-\nu_\tau$, or by improving inclusive searches~\cite{Heeck:2019kgr}.

\section*{Acknowledgements}
We thank Michael Schmidt and Sergey Syritsyn for explanations of the sign in Eq.~\eqref{eq:mapping} and  Ian Shoemaker for discussions. 
This work was supported in part by the National Science Foundation under Grant PHY-2210428 and a 4-VA at UVA Collaborative Research Grant.

\appendix

\section{Chiral perturbation theory}
\label{sec:chiralpt_appendix}

In this appendix we give a quick introduction to chiral perturbation theory to define our notation. We follow closely the derivation of Refs.~\cite{Claudson:1981gh,Chadha:1983sj,Chadha:1983mh,Nath:2006ut} but in a more explicit fashion that covers more operators.
Vector mesons can be included as in Ref.~\cite{Kaymakcalan:1983uc}, but since their pole diagrams typically dominate~\cite{Berezinsky:1981qb}, their branching ratios are suppressed compared to the pseudo-scalar mesons and depend on the same linear combinations of operators. 

Chiral perturbation theory provides an effective Lagrangian for the lightest hadrons -- those composed of up, down, and strange quarks -- by utilizing the approximate symmetries of the QCD Lagrangian, notably the global $SU(3)_{L} \times SU(3)_{R}$ symmetry under which $q_L\equiv (u_L,d_L,s_L) \sim (\vec{3},\vec{1})$ and $q_R\equiv (u_R,d_R,s_R) \sim (\vec{1},\vec{3})$. Together with the QCD gauge symmetry $SU(3)_C$ and the global Lorentz symmetry, we can then decompose products of quark fields such as $q_L q_L q_L$ into irreducible representations. If we restrict ourselves to color-singlet spin-$\tfrac12$ combinations -- which we can later identify with spin-$\tfrac12$ baryons -- the products of quarks are forced to form  an $(\vec{8},\vec{1})$ under $SU(3)_{L} \times SU(3)_{R}$, which can be explicitly written as a traceless $SU(3)_{L}$ matrix
\begin{align}
(q_L q_L q_L)_{(\vec{8},\vec{1})} = 
\matrixx{
(\overline{d}^{C}_{L} s_{L})\, u_{L}
&
(\overline{s}^{C}_{L} u_{L})\, u_{L}
&
(\overline{u}^{C}_{L} d_{L})\, u_{L} \\
(\overline{d}^{C}_{L} s_{L})\, d_{L}
&
(\overline{s}^{C}_{L} u_{L})\, d_{L}
&
(\overline{u}^{C}_{L} d_{L})\, d_{L} \\
(\overline{d}^{C}_{L} s_{L})\, s_{L}
&
(\overline{s}^{C}_{L} u_{L})\, s_{L}
&
(\overline{u}^{C}_{L} d_{L})\, s_{L}
} ,
\label{eq:qLqLqL_matrix}
\end{align}
where $(\overline{a}^{C}_{L} b_{L})\, c_{L} = - (\overline{b}^{C}_{L} a_{L})\, c_{L} \equiv \epsilon^{\alpha\beta\gamma} (\overline{a}^{C}_{L,\alpha} b_{L,\beta})\, c_{L,\gamma}$ with the $SU(3)_C$ Levi-Civita symbol $\epsilon^{\alpha\beta\gamma}$ and the two spinors in parenthesis form a Lorentz scalar.
Notice that the diagonal entries can be written in different ways upon using the vanishing $(\vec{1},\vec{1})$ trace
\begin{align}
(\overline{d}^{C}_{L} s_{L})\, u_{L} + (\overline{s}^{C}_{L} u_{L})\, d_{L} + (\overline{u}^{C}_{L} d_{L})\, s_{L} = 0\,.
\label{eq:qLqLqL_trace}
\end{align}
In complete analogy, we can obtain the $SU(3)_R$ matrix for the non-vanishing color-singlet spin-$\tfrac12$ piece $(q_R q_R q_R)_{(\vec{1},\vec{8})}$ by simply replacing $L \to R$ in Eq.~\eqref{eq:qLqLqL_matrix}.

The product $q_R q_R q_L$ can be treated similarly, now the color-singlet spin-$\tfrac12$ requirement projects out the $(\vec{3},\bar{\vec{3}})$ representation of $SU(3)_{L} \times SU(3)_{R}$, which in matrix notation reads
\begin{align}
(q_R q_R q_L)_{(\vec{3},\bar{\vec{3}})} = 
\matrixx{
(\overline{d}^{C}_{R} s_{R})\, u_{L}
&
(\overline{s}^{C}_{R} u_{R})\, u_{L}
&
(\overline{u}^{C}_{R} d_{R})\, u_{L} \\
(\overline{d}^{C}_{R} s_{R})\, d_{L}
&
(\overline{s}^{C}_{R} u_{R})\, d_{L}
&
(\overline{u}^{C}_{R} d_{R})\, d_{L} \\
(\overline{d}^{C}_{R} s_{R})\, s_{L}
&
(\overline{s}^{C}_{R} u_{R})\, s_{L}
&
(\overline{u}^{C}_{R} d_{R})\, s_{L}
} ,
\label{eq:qRqRqL_matrix}
\end{align}
with non-vanishing trace. 
$(q_L q_L q_R)_{(\bar{\vec{3}},\vec{3})}$ can be obtained from the above by interchanging $L$ and $R$. This concludes the representation theory of  color-singlet spin-$\tfrac12$ three-quark operators under the approximate global $SU(3)_{L} \times SU(3)_{R}$ symmetry. The components of these matrices show up in the $\Delta B \neq 0$ operators discussed in the main text and need to be mapped onto hadron operators with the same transformation properties. To this end we define the spin-$\tfrac12$ baryon matrix $B$ as well as the pseudo-scalar meson matrix $M$ and its logarithmically related matrix $\xi$ as
\begin{align}
B &= \begin{pmatrix}
        \sqrt{\frac{1}{2}}\Sigma^{0} + \sqrt{\frac{1}{6}} \Lambda^{0} & \Sigma^{+} & p \\
        \Sigma^{-} & -\sqrt{\frac{1}{2}}\Sigma^{0} + \sqrt{\frac{1}{6}} \Lambda^{0} &  n \\
        \Xi^{-} & \Xi^{0} & - \sqrt{\frac{2}{3}} \Lambda^{0}
        \end{pmatrix} ,\\
M &= \begin{pmatrix}
            \sqrt{\frac{1}{2}}\pi^{0} + \sqrt{\frac{1}{6}} \eta^{0} & \pi^{+} & K^{+} \\
        \pi^{-} & -\sqrt{\frac{1}{2}}\pi^{0} + \sqrt{\frac{1}{6}} \eta^{0} & K^{0} \\
        K^{-} & \overline{K}^{0} & - \sqrt{\frac{2}{3}} \eta^{0}
        \end{pmatrix}
        \equiv -\ii f_\pi \log \xi \,.
\end{align}
With these, we can construct products that have the same quantum numbers and transformation properties as our $qqq$ operators, specifically
\begin{align}
\xi B_L \xi \sim (\vec{3},\bar{\vec{3}})\,, && \xi^\dagger B_R \xi^\dagger \sim (\bar{\vec{3}},\vec{3})\,, && 
\xi B_L \xi^\dagger \sim (\vec{8},\vec{1})\,, && \xi^\dagger B_R \xi \sim (\vec{1},\vec{8})
\end{align}
under $SU(3)_{L} \times SU(3)_{R}$, where $B_{L,R}$ are the chiral components of $B$. These so-constructed matrices should then be proportional to the $qqq$ matrices with the same quantum numbers found above.
The proportionality factors contain the details of how quarks are confined into hadrons~\cite{Brodsky:1983st} through $SU(3)_C$ and can be obtained via lattice QCD~\cite{Yoo:2021gql}; since QCD conserves parity, only two are independent~\cite{Yoo:2021gql,Beneito:2023xbk},
\begin{align}
\begin{split}
(q_R q_R q_L)_{(\vec{3},\bar{\vec{3}})} &= \alpha\, \xi B_L \xi\,, \qquad (q_L q_L q_R)_{(\bar{\vec{3}},\vec{3})} = -\alpha\,  \xi^\dagger B_R \xi^\dagger \,, \\ 
(q_L q_L q_L)_{(\vec{8},\vec{1})} &= \beta\,\xi B_L \xi^\dagger\,, \quad\ \, (q_R q_R q_R)_{(\vec{1},\vec{8})} =  -\beta\,\xi^\dagger B_R \xi\,,
\end{split}
\label{eq:mapping}
\end{align}
and even those two are approximately connected via $\beta \simeq - \alpha$~\cite{Brodsky:1983st,Gavela:1988cp,Yoo:2021gql}. 
The matrix equations in Eq.~\eqref{eq:mapping} provide us with a dictionary for replacing three-quark operators with hadron operators, at least for the three lightest quarks and as a perturbation expansion in large pion-decay constant $f_\pi$. Together with the baryon-number-conserving interactions of the familiar ChPT Lagrangian~\cite{Chadha:1983sj} we can then calculate the effects of our $\Delta B$ operators in low-energy hadronic systems.
The $SU(3)_{L} \times SU(3)_{R}$ symmetry is of course only approximate in nature, so our Lagrangian should be supplemented with breaking-term corrections. Here, we only implement the most basic $SU(3)_{L} \times SU(3)_{R}$ breaking by giving the hadrons their measured non-degenerate masses.

\pagebreak
\bibliographystyle{JHEP}
\bibliography{BIB.bib}

\end{document}